\documentclass[final,5p,times,twocolumn]{elsarticle}

\usepackage{lipsum}
\usepackage{lineno,hyperref}
\usepackage{setspace}
\modulolinenumbers[1]

\journal{Journal of \LaTeX\ Templates}

\begin{document}

\begin{frontmatter}

\title{From vertex detectors to inner trackers with CMOS pixel sensors}

\author[StrasbourgUniversity]{A. Besson}

\author[IPHC]{A. P\'erez P\'erez\corref{Correspondingauthor}}
\cortext[Correspondingauthor]{Speaker}
\ead{Luis\_ Alejandro.Perez\_ Perez@iphc.cnrs.fr}

\author[LNF]{E. Spiriti}

\author[StrasbourgUniversity]{J. Baudot}
\author[IPHC]{G. Claus}
\author[IPHC]{M. Goffe}
\author[IPHC]{M. Winter}

\address[StrasbourgUniversity]{Universit\'e de Strasbourg, 4 rue Blaise Pascal, 67081 Strasbourg, France}
\address[IPHC]{IPHC-CNRS, 23 rue du loess, BP28, 67037 Strasbourg, France}
\address[LNF]{Laboratori Nazionali di Frascati, Italy}

\begin{abstract}
The use of CMOS Pixel Sensors (CPS) for high resolution and low material vertex detectors has been validated with 
the 2014 and 2015 physics runs of the STAR-PXL detector at RHIC/BNL. This opens the door to the use of CPS for 
inner tracking devices, with 10-100 times larger sensitive area, which require therefore a sensor design privileging 
power saving, response uniformity and robustness. The $350~{\rm nm}$ CMOS technology used for the STAR-PXL sensors
was considered as too poorly suited to upcoming applications like the upgraded ALICE Inner Tracking System (ITS), 
which requires sensors with one order of magnitude improvement on readout speed and improved radiation tolerance. 
This triggered the exploration of a deeper sub-micron CMOS technology, Tower-Jazz $180~{\rm nm}$, for the design of 
a CPS well adapted for the new ALICE-ITS running conditions. This paper reports the R\&D results for the conception 
of a CPS well adapted for the ALICE-ITS.
\end{abstract}


\end{frontmatter}


\section{Introduction}
\label{sec-intro}

\begin{figure}[htbp]
\begin{center}
\includegraphics*[width=4.0cm, height=2.5cm]{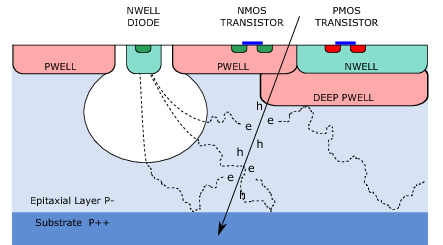}
\caption{Schematic cross-section of a CMOS pixel sensor.}
\label{figCMOSsensor}
\end{center}
\end{figure}

CPS integrate on the same silicon substrate the sensing elements and the front-end and readout 
circuitry (c.f. Fig.~\ref{figCMOSsensor}). Impinging charged particles create electron-hole pairs in a moderately 
P-doped epitaxial layer located on top of a highly P-doped (P++) wafer substrate and below some highly doped P-well (P++)
implants of the front-end circuitry. The generated electrons are collected on a N-well implanted on top of the epitaxial 
layer. In conventional CPS, the epitaxial layer is not fully depleted, and the electrons move mainly by thermal 
diffusion. However, they are deflected by build-in voltages of the P-epi/P++ interfaces which somewhat guide them toward 
the N-well/P-epi collection diode. Once collected, the charge is stored in the diode-parasitic capacitance, which voltage 
drop is amplified by a low-noise in-pixel amplifier.

CPS feature the possibility of fine pixel pitch (down to $10~{\rm \mu m}$) providing a very good spatial resolution (typically 
a few ${\rm \mu m}$). The very thin epitaxial layer ($10 - 40~{\rm \mu m}$) allows to thin the sensor down to $50~{\rm \mu m}$, 
which turns into an exceptionally small material budget. Furthermore, the sensors can be operated at room temperature, avoiding 
to use complicated cooling systems which additionally contribute to the material budget.

\begin{table}[]
\centering
\begin{TableSize}
\caption{Sensors design goals of STAR-PXL (operational) and new ALICE-ITS inner (ITS-in) and outer (ITS-out) layers. 
$\sigma_{\rm sp}$ refers to the spatial resolution, and TID and NIEL to the ionizing and non-ionizing doses, respectively.}
\label{tabSensorReq}
\begin{tabular}{l|c|c|c}
\hline
\hline
                                    & STAR-PXL & ITS (in) & ITS (out)  \\
\hline
$\sigma_{\rm sp}$ $[{\rm \mu m}]$   & $< 4$    & $< 5$    & $< 10$     \\
$t_{\rm r.o.}$ $[{\rm \mu s}]$      & $185.5$  & $30$     & $30$       \\
TID $[{\rm MRad}]$                  & $0.15$   & $2.70$   & $0.10$     \\
NIEL  $[{\rm 10^{12}~n_{eq}/cm^2}]$ & $3$      & $17$     & $1$        \\
$T_{\rm operation}$ $[^o{\rm C}]$   & $35$     & $30$     & $30$       \\
Power $[{\rm mW/cm^2}]$             & $160$    & $< 300$  & $< 100$    \\
Surface to cover $[{\rm m^2}]$      & $0.15$   & $0.17$   & $> 10$     \\
\hline
\hline
\end{tabular}
\end{TableSize}
\end{table}

A competitive tolerance to non-ionizing radiation (up to $10^{14}~{\rm n_{eq}/cm^2}$) was achieved when CMOS-processes with lightly-doped 
(so-called high-resistivity) epitaxial layer became available. This improved significantly the depletion depth up to several ${\rm \mu m}$, 
dramatically accelerating the charge collection, an thus reducing the transit time around trapping-defects in the bulk generated by non-ionizing 
radiation.

\begin{table*}[]
\centering
\begin{TableSize}
\caption{Properties (design goals for MISTRAL-O) of sensors discussed in this paper.}
\label{tabSensorsProperties}
\begin{tabular}{l|c|c|c|c}
\hline
\hline
                                   & MIMOSA-28          & FSBB-M0         & Mi-22THRb                        & MISTRAL-O         \\
\hline
CMOS process                       & $350~{\rm nm}$    & $180~{\rm nm}$   & $180~{\rm nm}$                   & $180~{\rm nm}$    \\
Pixels (col.$\times$row)           & $960\times928$    & $416\times416$   & $64\times64$                     & $832\times208$    \\
Pixel pitch [${\rm \mu m^2}$]      & $20.7\times20.7$  & $22\times33$     & $36\times62.5$ or $39\times50.8$ & $36\times65$      \\
Sensitive area [${\rm mm^2}$]      & $19.2\times19.9$  & $13.7\times9.2$  & $8 - 9$                          & $13.5\times30.0$  \\
$\sigma_{\rm sp}$ [${\rm \mu m}$]  & $\gtrsim 3.6$     & $\gtrsim 4.5$    & $\sim 10$                        & $\sim 10$         \\
$t_{\rm r.o.}$ [${\rm \mu s}$]     & $185.5$           & $41.6$           & $\sim 5$                         & $20.8$            \\
TID [${\rm MRad}$]                 & $> 0.15$          & $ > 1$           & $> 0.15$                         & $> 0.15$          \\
NIEL [$10^{12}{\rm n_{eq}/cm^2}$]  & $> 3$             & $ > 10$          & $> 1$                            & $> 1$             \\
Power [${\rm mW/cm^2}$]            & $160$             & $< 160$          & N/A                              & $\lesssim 80$     \\
Pads over pixels                   & No                & No               & Yes                              & Yes               \\
On-chip sparsification             & 1D                & 2D               & None                             & 2D                \\
\hline
\hline
\end{tabular}
\end{TableSize}
\end{table*}

The sensor's readout speed depends on the on-chip data processing circuitry. Initially, the signal from the in-pixel 
amplifier was multiplexed to a common analog readout bus and sent out for further processing, giving a few ${\rm ms}$ 
readout time ($t_{\rm r.o.}$) for a sensor of larger sensitive area. A factor of 1000 reduction in $t_{\rm r.o.}$ was 
obtained by reading in succession rows of the pixel matrix, allowing each pixel to send in parallel their signal to the 
column end, the so-called rolling shutter architecture. This data is then processed trough discriminators ending each 
column for digitization and then further processed by an on-chip data sparsification circuit. This allowed parallel column 
readout and reduced transmission band-width, shortening the $t_{\rm r.o.}$ to about $100~{\rm \mu s}$.

All these developments allowed the conception of a CPS, called MIMOSA-28~\cite{Mi28_ref} (c.f. table~\ref{tabSensorsProperties}), 
suited to the STAR-PXL detector~\cite{STAR_PXL_Dect} operated at RHIC/BNL, the first vertex detector based on the CPS technology. 
The STAR-PXL has successfully participated to two data-taking campaigns and is currently in operation. It has proven to be reliable 
and to deliver the expected added value for the STAR physics program. The 400 sensors composing the PXL have allowed accumulating 
a sizable amount of experience with these devices.

The performances reached by the MIMOSA-28 sensor are not suited to match some of the more demanding requirements of the new 
ALICE-ITS~\cite{ALICE_ITS_TDR} (c.f. table~\ref{tabSensorReq}), mainly in terms of $t_{\rm r.o.}$ and power consumption 
(outer layers), and to a lesser extent in terms of radiation tolerance~\footnote{The ALICE-ITS doses in table~\ref{tabSensorReq} 
are the ones reported in the TDR, but more detailed studies have shown that significantly lower doses need to be tolerated.} 
(inner layers). This triggered the exploration of a new $180~{\rm nm}$ CMOS process with several advantages to overcome the limitations 
of the AMS $350~{\rm nm}$ CMOS process used for MIMOSA-28. This document 
describes the features of the new CMOS process and its advantages as compared to elder ones. Furthermore, current status 
of the R\&D for the conception of a sensor adapted for the ALICE-ITS outer layers is also presented.

\section{ALICE-ITS upgrade: R\&D of sensors for outer layers}
\label{sec-RandD}

To overcome the limitations of the MIMOSA-28 sensor for the ALICE-ITS application, it 
was decided to migrate to the novel Tower-Jazz $180~{\rm nm}$ CMOS process. One of its aspects is the smaller feature 
size, improving the $t_{\rm r.o.}$ (higher integration density) and tolerance to ionizing radiation. Furthermore, the process 
grants thicker epitaxial layers with higher resistivity, improving the signal-to-noise ratio and tolerance to 
non-ionizing radiation. Finally, the new technology concedes for deep P-well implants which shield the N-well hosting PMOS 
transistors, preventing parasitic charge collection. This allows both PMOS and NMOS transistors to be used 
inside pixels, permitting to implement an in-pixel discriminator.

All these features widen the choice of the readout architecture strategy. An asynchronous readout similar to the 
one used for the hybrid pixel sensors~\cite{FEI4_ref} is pursued via the ALPIDE design~\cite{ALPIDE_ref}, which is the most promising approach 
to equip the ALICE-ITS inner layers. This approach comes with certain risks of missing the project's tight schedule, as it 
requires building and testing a complex pixel matrix with several new features to be validated. In order to be on 
schedule it was decided to also follow a more conservative approach, the MISTRAL-O chip, using the validated rolling shutter readout of the MIMOSA-28 
sensor. MISTRAL-O was intended to instrument the ALICE-ITS outer layers, which feature two orders of magnitude higher surface to cover, 
thus requiring special attention on power consumption, response uniformity and robustness.

MISTRAL-O needs to be about 10 times faster than MIMOSA-28 together with twice less power consumption (c.f. table~\ref{tabSensorReq}). 
It has as well to be pin-to-pin compatible with the ALPIDE design, which includes pads over the pixel matrix suited to laser soldering. 
Furthermore, the sensor's slow control and digital logic need to be adapted to ALPIDE's standards.

\subsection{Rolling shutter readout in the novel CMOS-process}
\label{subsec-readout}

As a first step for validating the new $180~{\rm nm}$ CMOS-process, a full scale prototype called FSBB-M0 was built early in 2015 (c.f. 
table~\ref{tabSensorsProperties}). The sensor includes a very similar analog front-end and digital readout chain as the MIMOSA-28 sensor
with a $t_{\rm r.o.}$ reduced by a factor of $\sim 4$. This was achieved by reducing the number of pixels in a column to be readout 
($928 \to 416$) and increasing the size of the pixel long the column ($20.7 \to 33~{\rm \mu m}$). Furthermore, the rolling shutter readout 
mode addresses simultaneously all pixels belonging to a pair of neighboring rows. Therefore, each discriminator addresses 208 pixels 
instead of 928, with a proportionnally reduced $t_{\rm r.o.}$. The sensing nodes are staggered in order to maximize the uniformity of the 
sensing node density and to alleviate the spatial resolution asymmetry consecutive to the pixel's rectangular shape. Moreover, the sparsification 
circuit allows to find clusters of pixels in 2-dimensional windows instead of the 1-dimensional ones of MIMOSA-28, which had to be merged 
off-line.

\begin{figure}[htbp]
\begin{center}
\includegraphics*[width=7.0cm, height=4.0cm]{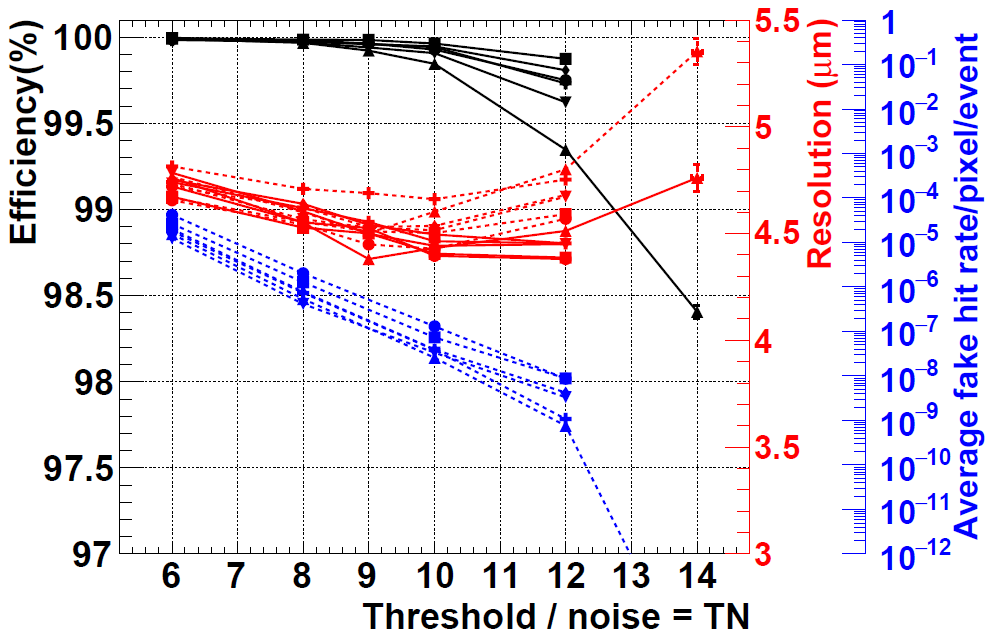}
\caption{Detection efficiency (black), dark occupancy (dotted-blue) and spatial resolution along short (solid-red) and long (dotted-red) pixel 
sides as a function of the discriminator threshold (in multiples of thermal noise) of the FSBB-M0 sensor. The performances of 6 sensors are superimposed.
$T_{\rm operation} = {\rm 30^oC}$.}
\label{fig-FSBBbis_nonIrradiated}
\end{center}
\end{figure}

Several FSBB-M0 sensors were first studied in the laboratory, where their temporal and fixed pattern noise performances were assessed over a wide 
range of positive temperatures. These measurements were performed before and after irradiation with an X-Ray source (up to $1.6~{\rm MRad}$) and 
with $1~{\rm MeV}$ neutrons (up to $10^{13}~{\rm n_{eq}/cm^2}$), up to doses relevant for the ALICE-ITS inner layers (c.f. table~\ref{tabSensorReq}). The 
sensors were next tested with particle beams at the CERN-SPS with negatively charger pions of $\sim 120~{\rm GeV/c}$, and at DESY 
with electrons of $3 - 6~{\rm GeV/c}$. Each set-up was made of six FSBB-M0 sensors operated simultaneously on beam. The performances of each 
sensor were assessed by considering it as a Device Under Test (DUT) and the rest as reference planes for track reconstruction (telescope). The 
measurements showed satisfactory results, ilustrated by Fig.~\ref{fig-FSBBbis_nonIrradiated}. A nearly $100\%$ detection efficiency and a dark 
occupancy of $< 10^{-5}$ was observed for all non-irradiated sensors tested. The dark occupancy could be reduced by an order of magnitude by 
masking a small fraction ($< 0.5\%$) of the noisiest pixels, with negligible impact on the detection efficiency. Satisfactory 
detection efficiency ($> 99\%$) was also obtained for sensors irradiated at doses relevant for the ALICE-ITS inner layers, and dark occupancies of 
$< 10^{-5}$ were obtained by masking  the $0.5\%$ noisiest pixels.

\begin{figure}[htbp]
\begin{center}
\includegraphics*[width=3.0cm, height=3.0cm]{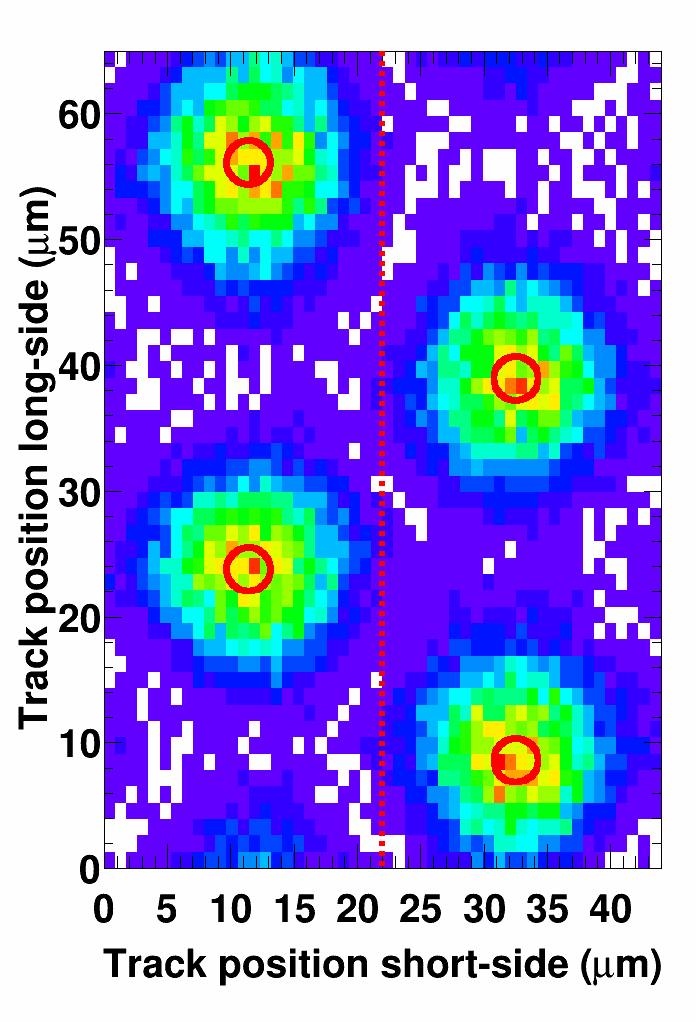}
\includegraphics*[width=3.0cm, height=3.0cm]{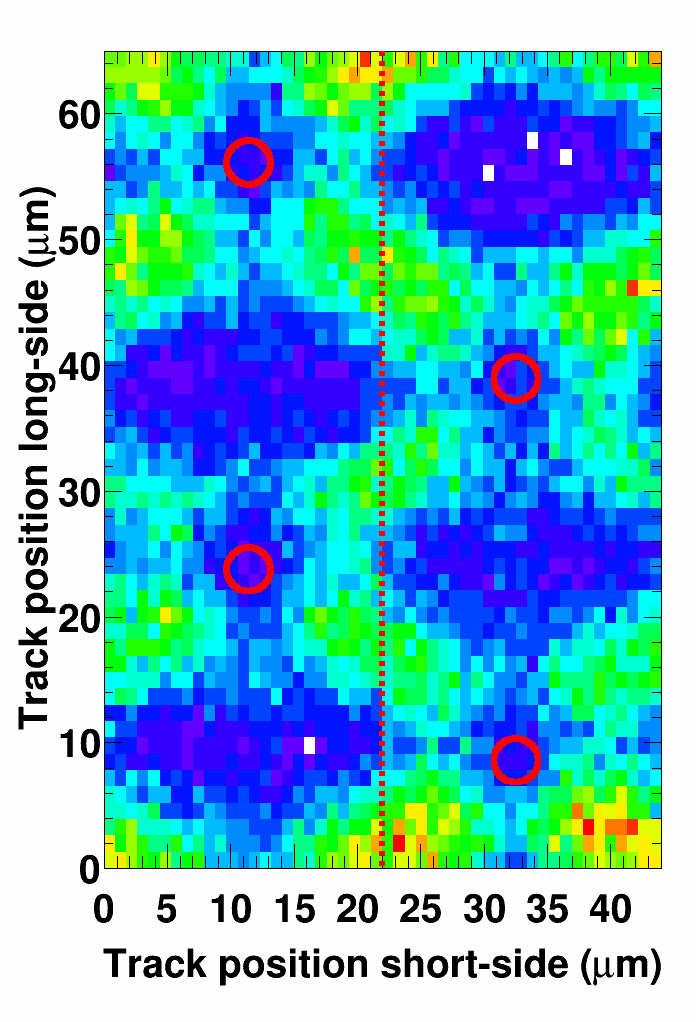}
\includegraphics*[width=3.0cm, height=3.0cm]{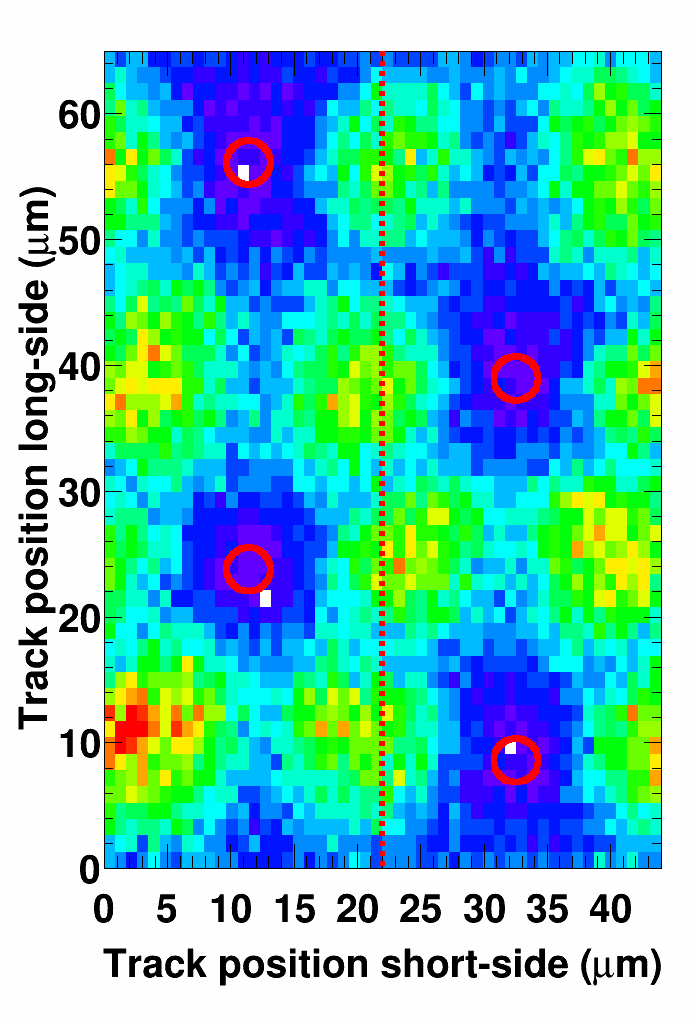}
\includegraphics*[width=3.0cm, height=3.0cm]{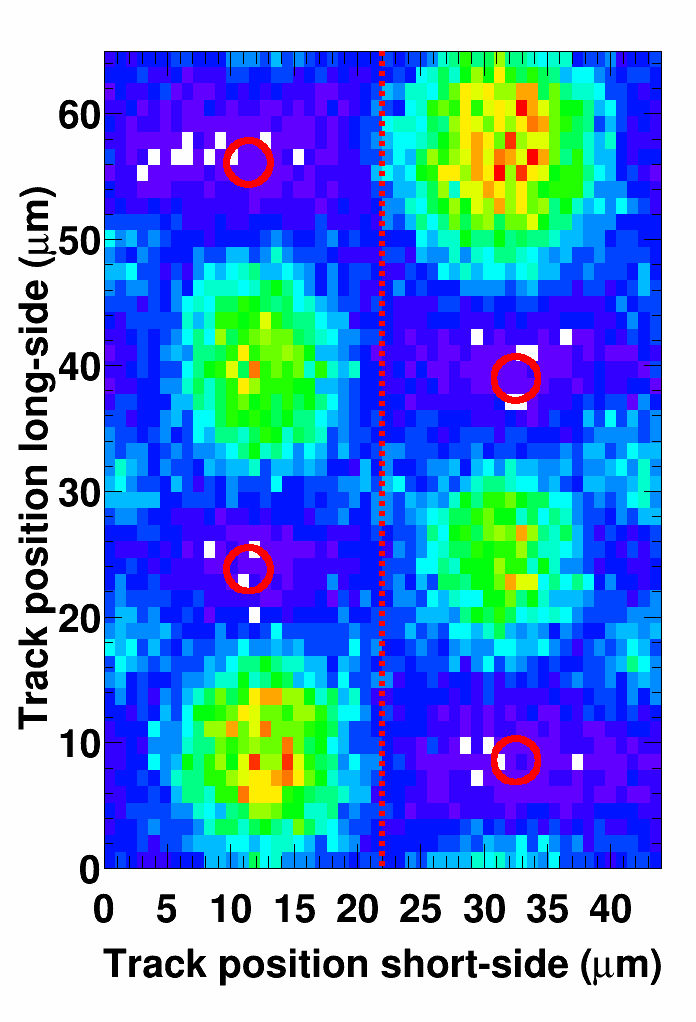}
\caption{Track impact position at DUT with respect the closest collection diode (red circles). The four near-most 
collection diodes are shown. The vertical red-dotted line shows the boundary between two columns of pixels. 
The plots from left to right and top to bottom corresponds to hit pixel multiplicities from 1 to 4, respectively. 
The telescope pointing resolution is $2.2~{\rm \mu m}$.}
\label{fig-FSBBbis_TrkPositionWrtDiode_vs_Mult}
\end{center}
\end{figure}

In order to estimate the telescope pointing resolution ($\sigma_{\rm Tel}$) at the DUT position, the experimental set-up was simulated with a 
Geant4-based tool. A $\sigma_{\rm Tel}$ from $2.2$ to $4.4~{\rm \mu m}$ ($2.7$ to $5.5~{\rm \mu m}$), was obtained at the CERN-SPS (DESY) beam 
test facility. The low (high) values of $\sigma_{\rm Tel}$ correspond to the DUT located at the center (ends) of the sensors layout. 
This allowed to estimate the DUT spatial resolution ($\sigma_{\rm sp}$) by subtracting the $\sigma_{\rm Tel}$ from the track-hit residue. As 
shown in Fig.~\ref{fig-FSBBbis_nonIrradiated}, a $\sigma_{\rm sp}$ of $\sim 4.5~{\rm \mu m}$ was obtained for all six sensors tested, with 
a slightly higher value along the long-side of the rectangular pixel ($22\times33~{\rm \mu m^2}$). It is substantially better than one would 
expect for the digital readout according to the $\sigma_{\rm digital} = {\rm pitch}/\sqrt{12}$ rule. This follows from charge charing, giving 
a hit pixel multiplicity of about 3 in average, thereby providing a substantially better $\sigma_{\rm sp}$ based on the center-of-gravity of 
the cluster of pixels. The sensitivity of $\sigma_{\rm sp}$ to charge sharing is shown in Fig.~\ref{fig-FSBBbis_TrkPositionWrtDiode_vs_Mult}, 
which displays, for different hit pixel multiplicities, the tracks impact position with respect to the closest collection diode. 
One can appreciate the regions with respect to the collection diodes where an impinging particle needs to hit the sensor in order 
to produce a given pixel multiplicity. The sizes of those regions give a measurement of the spatial resolution, which vary smoothly from one 
hit pixel multiplicity to the other. Even if a subset of about $15\%$ of the hits show only one pixel, they exhibit a $\sigma_{\rm sp}$ 
much better than $\sigma_{\rm digital}$ in both pixel directions. Such single pixel clusters are formed exclusively by particles impinging the 
pixel near the collection diode (top-left plot of Fig.~\ref{fig-FSBBbis_TrkPositionWrtDiode_vs_Mult}), otherwise clusters include systematically 
more than one pixel.

\subsection{Large pixel design and validation}
\label{subsec-LargePixel}

The observed detection performances of the FSBB-M0 prototype validated the implementation of the rolling shutter architecture with 
the new $180~{\rm nm}$ CMOS-process. The next step consisted in addressing the ALICE-ITS requirements by achieving nearly twice 
less readout time and power consumption. A large fraction of the power consumption 
is due to the in-pixel circuitry. Power savings can then be achieved by minimizing the number of pixels per sensor, i.e. by increasing 
their dimensions. This approach exploited the alleviated spatial resolution requirement for the outer layers as compared to the inner 
layers (c.f. table~\ref{tabSensorReq}). Furthermore, the use of large pixels elongated in the row 
direction further downscales the number of pixels per discriminator and the number of columns per surface unit, reducing in this way 
the $t_{\rm r.o.}$ and the power consumption. The increase of the pixel surface has however drawbacks in terms of charge collection. 
The mean distance the signal charges cross until being collected increases as the collection diode density is reduced, increasing the 
probability for charge trapping by defects of the silicon crystalline structure. This probability is further enhanced after non-ionizing 
irradiation. However, this effect is foreseen to remain acceptable given the relaxed requirements in the ALICE-ITS outer layers.

CPS exploring large pixel detection properties (Mi-22THRb, c.f. table~\ref{tabSensorsProperties}) were manufactured in 2014 in the  
Tower-Jazz $180~{\rm nm}$ CMOS process featuring 6 metalization layers and deep P-wells implants. Two different pixel dimensions of 
$36\times62.5~{\rm \mu m}^2$ and $39\times50.8~{\rm \mu m}^2$ were 
implemented. Each pixel incorporates a sensing node connected to a pre-amplifier featuring a feedback loop ended with a forward biased 
diode compensating the leakage current delivered by the sensing node and fixing its depletion voltage. The preamplifier is followed by 
a clamping circuit subtracting the pixel pedestal and connected to a discriminator located at the column end. The sensing nodes are 
staggered in order to maximize the uniformity of the sensing node density and to alleviate the spatial resolution asymmetry consecutive 
to the pixel's rectangular shape. The matrix, made of 64 columns of 64 pixels, is read out in rolling shutter mode addressing simultaneously 
all pixels belonging to a pair of neighboring rows. 8 columns feature (discriminator free) analog output for pixel circuitry assessment 
purposes. The $t_{\rm r.o.}$ of the sensor is $\sim 5~{\rm \mu s}$.

The relatively large pixel size allows to integrate in-pixel circuitry for noisy pixel masking. Furthermore, the dimensions enable to 
concentrate the in-pixel circuitry within 3 metalization layers only. This allows integrating connection pads over the pixel array, 
using the two top metal layers. In the MIMOSA-28 sensor, a MIM (Metal-Insulator-Metal) clamping capacitor was implemented using the two 
top metalization layers. The connection pads over the pixel array forced to explore new alternatives of the clamping capacitor implementation.

The prototyping of the large pixel concept addressed in particular the detection efficiency, the spatial resolution, the radiation tolerance 
and the in-pixel circuitry with a new clamping design. For this purpose, several pixel variants were designed and implemented in 8 different 
pixel arrays, including a reference one based on 6 metalization layers.

\begin{figure}[htbp]
\begin{center}
\includegraphics*[width=7.0cm, height=3.5cm]{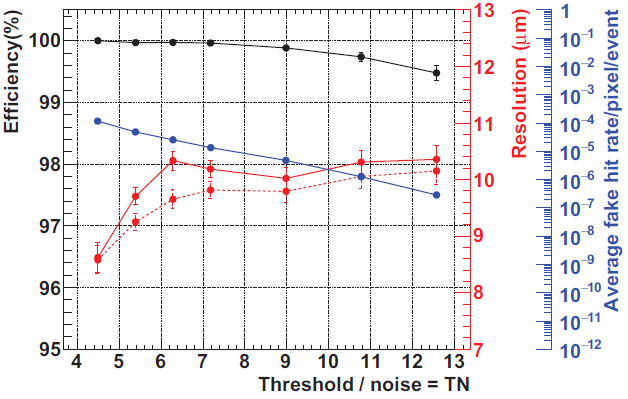}
\caption{Detection efficiency (black), dark occupancy (blue) and spatial resolution along short (solid-red) and long (dotted-red) pixel sides 
as a function of the discriminator threshold (in multiples of thermal noise) of the Mi-22THRb sensor with pixels dimensions of $39\times50.8~{\rm \mu m^2}$. 
$T_{\rm operation} = {\rm 30^oC}$.}
\label{figMi22_BTF_LNF}
\end{center}
\end{figure}

The sensors were first studied in the laboratory, where their temporal and fixed pattern noise performances were assessed over a wide range of 
positive temperatures and where their response to an $^{55}{\rm Fe}$ source were studied and used to calibrate the charge-to-voltage conversion 
gain; these measurements were performed before and after irradiation with an X-Ray source (up to $150~{\rm kRad}$) and with $1~{\rm MeV}$ 
neutrons (up to $1.5\times10^{12}~{\rm n_{eq}/cm^2}$), up to doses relevant for the ALICE-ITS outer layers. The sensors were next tested on an 
electron beam of $450~{\rm MeV/c}$ at the Frascati beam test facility (LNF/BTF)~\cite{BTF_LNF}. A telescope composed of analog output 
sensors (MIMOSA-18~\cite{Mi18_ref}) featuring a $1-2~{\rm \mu m}$ spatial resolution were used to predict the beam particle impact position at 
the DUT (Mi-22THRb sensors) location. The telescope was simulated in detail, accounting for multiple scattering. The predicted telescope pointing 
resolution amounts to $5 - 6~{\rm \mu m}$. Fig.~\ref{figMi22_BTF_LNF} shows a representative result of this beam test. For both pixel dimensions, 
a nearly $100\%$ detection efficiency was observed for non-irradiated sensors. Satisfactory detection efficiency ($> 99\%$) was also obtained for sensors 
irradiated at doses relevant of the ALICE-ITS outer layers. Moreover, a spatial resolution of $\sim 10~{\rm \mu m}$ in both directions was found 
for all pixel variants tested. All these results comply with all ambitioned performances of the large pixel design well adapted for the ALICE-ITS 
outer layers (c.f. table~\ref{tabSensorReq}).

\section{Summary and outlook}
\label{sec-Summary}

This paper summarizes the R\&D results for the design of a CPS well adapted to the ALICE-ITS outer layers. Two prototypes were 
fabricated and fully tested. The first one addressed the validation of the rolling shutter readout architecture implemented in 
the new Tower-Jazz $180~{\rm nm}$ CMOS-process, with nearly four times shorter $t_{\rm r.o.}$ than the MIMOSA-28 sensor equipping the STAR-PXL. 
The second prototype addressed a large pixel design, including bonding pads over the pixel matrix, to further squeeze $t_{\rm r.o.}$ 
and the power consumption, while complying with $\sigma_{\rm sp}$ and radiation hardness requirements of the ALICE-ITS outer layers.
The results obtained validate the possibility to fabricate the MISTRAL-O sensor (c.f. table~\ref{tabSensorsProperties}), featuring 
$\sim 160{\rm k}$ pixels over $4~{\rm cm}^2$ of sensitive area, a $t_{\rm r.o.} \sim 21~{\rm \mu s}$ and dissipating less than 
$100~{\rm mW/cm}^2$, well within the target requirements (c.f. table~\ref{tabSensorReq}).

The results presented in this paper validate the Tower-Jazz $180~{\rm nm}$ CMOS technology, as well as the strategies for improving
the readout speed, power consumption and radiation tolerance for future applications. Among them the vertex detector (MVD) of the
forthcoming CBM experiment~\cite{CMB_MVD_ref} at FAIR/GSI and a vertex detector suited to experiments foreseen at the ILC~\cite{ILC_TDR}.


\section*{Acknowledgements}

We thank our colleagues C. Bedda and I. Ravasenga (INFN-Torino), and P. La Rocca and F. Riggi (INFN-Catania), 
for their contributions to the beam test of Mi-22THRb at LNF/BTF.



\end{document}